\documentclass[prl,twocolumn,preprintnumbers,amsmath,amssymb]{revtex4}% Physical Review B

\usepackage{graphicx}% Include figure files
\graphicspath{{fig/}}
\usepackage{dcolumn}% Align table columns on decimal point\varphi 
\usepackage{bm}% bold math
\usepackage{amsmath,amssymb}
\usepackage[T1]{fontenc}
\usepackage{textcomp}
\usepackage{color}
\begin{document}

\title{Magnonic Noise and Wiedemann-Franz Law
}

\author{Kouki Nakata,$^{1}$ Yuichi Ohnuma,$^{2}$ and Mamoru Matsuo$^{2}$}

\affiliation{$^1$Advanced Science Research Center, Japan Atomic Energy Agency, Tokai, Ibaraki 319-1195, Japan  \\
$^2$Kavli Institute for Theoretical Sciences, University of Chinese Academy of Sciences, Beijing, 100190, China
}

\date{\today}

\begin{abstract}
We theoretically establish mutual relations among magnetic momentum, heat, and fluctuations of propagating magnons in a ferromagnetic insulating junction in terms of noise and the bosonic Wiedemann-Franz (WF) law.
Using the Schwinger-Keldysh formalism, we calculate all transport coefficients of a noise spectrum for both magnonic spin and heat currents, and establish Onsager relations between the thermomagnetic currents and the zero-frequency noise.
%%%%%%%%%%%%%%%%%%%%%%%%
Making use of the magnonic WF law and the Seebeck coefficient in the low-temperature limit, we theoretically discover universal relations, i.e. being independent of material parameters, for both the nonequilibrium and equilibrium noise, and show that each noise is described solely in terms of thermal conductance. 
%being measurable with current device and measurement techniques.
%%%%%%%%%%%%%%%%%%%%%%%%
Finally, we introduce a magnonic spin-analog of the Fano factor, noise-to-current ratio, and demonstrate that the magnonic spin-Fano factor reduces to a universal value in the low-temperature limit and it remains valid even beyond a linear response regime.
%%%%%%%%%%%%%%%%%%%%%%%%
%Finally, we show that our theoretical predictions are within experimental reach with current device and measurement techniques.
\end{abstract}

%\pacs{75.30.Ds, 73.43.-f, 75.47.-m, 77.55.Nv, 85.75.-d,72.25.-b, 75.85.+t}

\maketitle

%%%%%%%%%%%%%%%%%
%\section{Introduction}
%\label{sec:Intro}
%%%%%%%%%%%%%%%%%
{\it{Introduction.$-$}}
%%%%%%%%%%%%
%%Meso;background
%%%%%%%%%%%%
Nonequilibrium noise under a high voltage bias, shot noise \cite{ShotNoise1918,NoiseRev}, arises from a nonequilibrium fluctuation of the electric current in mesoscopic systems, and provides abundant information about electron transport \cite{LandauerNoise,DLnoise6,DLnoise5,DLnoise4,DLnoise3,DLnoise2,DLnoise}.
Being the result of charge quantization, the shot noise allows to characterize the transport properties of individual electrons  (e.g., the Fano factor), and serves as a sensitive tool to probe an effective charge \cite{NoiseExp_e,NoiseExp_e2,NoiseExp_e3,NoiseExp_e4,NoiseExp_e5,NoiseExp_e6}.

%%%%%%%%%%%%%%%%%%%%%%%%
%%Spin current noise;background2, SaikinnnoSinten
%%%%%%%%%%%%%%%%%%%%%%%%
Despite the fact that most studies focus on electric conductors or semiconductors (e.g., quantum dot), these fundamental concepts also extend to spin degrees of freedom, namely, the noise of spin currents \cite{SpinNoiseMeasure,Kamra2,Kamra3,noise2018,STnoise}.
Making use of the inverse spin Hall effect (ISHE) \cite{ISHEadd,ishe,ISHEadd2} to convert spin currents into electric currents in a bilayer system formed by a normal metal and a ferromagnetic insulator (FI), 
Ref. \cite{SpinNoiseMeasure} reported measurements of a spin current noise as the ISHE-induced voltage noise.
%%%%%%%%%%%%%%%%%%%%
%%problem, motivation; maefuri
%%%%%%%%%%%%%%%%%%%%$\ddot{\rm{u}}$ 
However, the Onsager relations, being one of the key ingredients in the fluctuation-dissipation theorem \cite{FDT,FDT2,FDT3,FDT4}, between the spin current and the noise have not yet been explored; while the last decade has seen remarkable development of spin caloritronics \cite{spincal} (e.g., the observation of spin Seebeck \cite{uchidainsulator} and Peltier \cite{Peltier} effects) and the understanding of the conversion mechanism between spin and heat transport has been well-developed, the mutual relations among spin, heat, and fluctuations remain an open issue.

\begin{figure}[t]
\begin{center}
\includegraphics[width=8cm,clip]{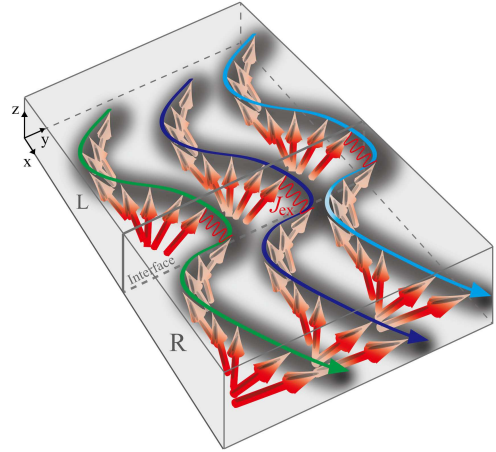}
\caption{Schematic illustration of the ferromagnetic insulating junction.
The boundary spins in the left (L) and right (R) FIs are relevant to the transport of magnons (i.e., quanta of spin-waves) with fluctuations across the junction interface, and they are weakly exchange-coupled with the strength $J_{\rm{ex}}$. 
}
\label{fig:system}
\end{center}
\end{figure}

%%%%%%%%%
%%purpose
%%%%%%%%%
In this paper, we provide a solution to this fundamental challenge in terms of the Wiedemann-Franz (WF) law \cite{WFgermany} for magnon transport \cite{magnonWF,KJD,KSJD,ReviewMagnon}, which characterizes the universal thermomagnetic properties of the magnonic spin and heat currents; 
establishing the Onsager relations between the magnonic currents \cite{spinwave,WeesNatPhys} and the noise, 
we develop the law further into the one for a noise spectrum, namely, the {\it{WF law for the magnonic noise}}.
We thus theoretically explore universal relations for both the nonequilibrium  and equilibrium noise, and demonstrate that a magnonic spin-analog of the Fano factor reduces to a universal value  even beyond a linear response regime.
%and it remains valid

%%%%%%%%%magnonWF\cite{ShotNoise1918}\cite{ThermalNoise,ThermalNoise2}
%%Remark
%%%%%%%%%
For this purpose, FIs are the best platform \cite{MagnonSpintronics,magnonics} since, in complete absence of any conducting elements, they host magnons which are chargeless bosonic quasiparticles with a magnetic dipole moment $g \mu_{\rm{B}} {\mathbf{e}}_z$  that serve as information carriers in Bohr magneton units $\mu_{\rm{B}}$.
Thereby we can extract the intrinsic properties of chargeless spin and heat transport with fluctuations.

%%%%%%%%%%%%%%%%%
%\section{System}
%\label{sec:system}
%%%%%%%%%%%%%%%%%
{\it{System.$-$}}
%%%%%%%%%%%%
We consider a magnetic junction formed by two FIs aligned along the $x$-direction, see Fig. \ref{fig:system}.
Assuming cubic lattices, the left FI is identical to the right one and each of the three-dimensional FIs can be described by a Heisenberg spin Hamiltonian in the presence of a Zeeman term.
Applying a strong magnetic field $ B_{\rm{L(R)}}  $ to the left (right) FI, we assume ferromagnetic order along the external magnetic field, defining the $z$-direction, and that the resulting Zeeman energy $ g \mu_{\rm{B}} B_{\rm{L(R)}}$ is the leading contribution to the Zeeman term.
The temperature of the left (right) FI is  $ T_{\rm{L(R)}}  $.
%%%%%%%%%%%%%%%%%%%%%%%%%%%%%%%%%
A finite overlap of the wave functions of the boundary spins ${\bf S}_{\Gamma_{\rm{L}}}$ and ${\bf S}_{\Gamma_{\rm{R}}}$ in the left and right FI, respectively, results in an exchange interaction, weakly coupling the two FIs, which may be described by the Hamiltonian 
${\cal{H}}_{\rm{ex}}  = -J_{\rm{ex}} \sum_{\langle \Gamma_{\rm{L}} \Gamma_{\rm{R}} \rangle} {\bf S}_{\Gamma_{\rm{L}}} \cdot {\bf S}_{\Gamma_{\rm{R}}}$,
where we assume that the exchange interaction $J_{\rm{ex}} >0 $ between the two FIs is weak compared with the one between the nearest neighbor spins in each FI. 
%%%%%%%%%%%%%%%%%%%%%%%
Considering large spins of length $S \gg  1$ and thereby performing a Holstein-Primakoff expansion \cite{HP} to leading order, the Hamiltonian reduces to
$ {\cal{H}}_{\rm{ex}} =  - J_{{\rm{ex}}} S \sum_{{\mathbf{k}}_{\perp }}  \sum_{k_x, k_x^{\prime}}
a_{{\rm{L}},  {\mathbf{k}}}  a^{\dagger }_{{\rm{R}}, {\mathbf{k}}^{\prime}} + {\rm{H. c.}}   $,
where the bosonic operator $a_{\rm{L/R}}^{\dagger}$ ($a_{\rm{L/R}}$) creates (annihilates) a boundary magnon at the left/right FI with the momentum
${\mathbf{k}}=(k_x, k_y, k_z)$ and ${\mathbf{k}}^{\prime}=(k_x^{\prime}, k_y, k_z)$, respectively, and where ${\mathbf{k}}_{\perp }=(0, k_y, k_z)  $.
%Since we assume a sharp junction interface, the $k_x$-momentum of propagating magnons across the interface is not conserved, while the perpendicular component ${\mathbf{k}}_{\perp }$ is conserved.
Throughout this paper, we focus on sufficiently low temperatures where effects of magnon-magnon and magnon-phonon interactions become negligibly small \cite{Tmagnonphonon,adachiphonon}.

%%%%%%%%%%%%%%%%%%%%
%\section{Magnonic noise}
%\label{sec:noise}
%%%%%%%%%%%%%%%%%%%%
{\it{Magnonic noise.$-$}}
%%%%%%%%%%%%%%%%%%%%
The tunneling Hamiltonian $ {\cal{H}}_{\rm{ex}}$ gives the time-evolution of both the magnon number operator and the energy operator for each FI,
and generates those magnonic currents across the junction interface.
The Heisenberg equation of motion provides the magnonic spin and heat current operators
$ \hat{{\cal{I}}}_{\rm{m}} (t)$ and  $ \hat{{\cal{I}}}_{Q} (t)$, respectively, 
which flow across the junction interface from the right FI to the left one \cite{PeltierOhnuma};
$ {\hat{{\cal{I}}}}_{\rm{m}}  (t) = - i  g \mu _{\rm{B}} (J_{\rm{ex}}S/\hbar ) \sum_{{\mathbf{k}}, k_x^{\prime}}
a_{{\rm{L}},{\mathbf{k}}}(t)  a^{\dagger }_{{\rm{R}},{\mathbf{k}}^{\prime}}(t) + {\rm{H. c.}}$,
%%%%%%%%%%%%%%%%%%%%%%%%%%%%%%%%
$ {\hat{{\cal{I}}}}_Q  (t) = J_{\rm{ex}}S  \sum_{{\mathbf{k}}, k_x^{\prime}}
 [\partial _{t}^{\rm{L}} a_{{\rm{L}},{\mathbf{k}}}(t)]  a^{\dagger }_{{\rm{R}},{\mathbf{k}}^{\prime}}(t) + {\rm{H. c.}}$,
 where $ \partial _{t}^{\rm{L}}  $ denotes the time-derivative which works on the magnon operators solely for the left FI.
%%%%%%%%%%%%%%%%%%%%%%%%%%%%%%%
%%%%%%%%%%%%%%%%%%%%%%%%%%%%%%%

Introducing spin and heat current-current correlation functions as the statistical average
$ {\cal{S}}_{\rm{m}}(t, t') \equiv  \langle  \{\hat{{\cal{I}}}_{\rm{m}}(t),  \hat{{\cal{I}}}_{\rm{m}}(t')  \} \rangle /2  =   \langle  \{\hat{{\cal{I}}}_{\rm{m}}(t) - \langle  \hat{{\cal{I}}}_{\rm{m}} \rangle,  \hat{{\cal{I}}}_{\rm{m}}(t') - \langle  \hat{{\cal{I}}}_{\rm{m}} \rangle  \} \rangle /2 + {\cal{O}} (J_{\rm{ex}}^4) $ and 
%%%%%%%%%%%
$ {\cal{S}}_Q (t, t') \equiv  \langle  \{\hat{{\cal{I}}}_Q(t),  \hat{{\cal{I}}}_Q(t')  \} \rangle /2  =   \langle  \{\hat{{\cal{I}}}_Q(t)- \langle  \hat{{\cal{I}}}_Q \rangle,  \hat{{\cal{I}}}_Q(t')- \langle  \hat{{\cal{I}}}_Q \rangle  \} \rangle /2 + {\cal{O}} (J_{\rm{ex}}^4) $, respectively,
we assume the steady state in terms of time \cite{haug,PeltierOhnuma}
 ${\cal{S}}_{{\rm{m}}(Q)}(t, t') = {\cal{S}}_{{\rm{m}}(Q)}(\delta {t}) $ and $\delta {t} \equiv  t-t'  $.
%%%%%%%%%%%%%%%%%%
Defining a noise spectrum for each magnonic current
$ {\cal{S}}_{{\rm{m}}(Q)}(\Omega ) \equiv  \int d  (\delta {t}) {\rm{e}}^{i \Omega  \delta {t} }   {\cal{S}}_{{\rm{m}}(Q)}(\delta {t})   $
and taking the dc-limit $   {\cal{S}}_{{\rm{m}}(Q)}(\Omega =0) \equiv    {\cal{S}}_{{\rm{m}}(Q)}$,
the magnonic spin and heat noise ${\cal{S}}_{{\rm{m}}}$ and ${\cal{S}}_Q$, respectively, are introduced.
%%%%%%%%%%%%%%%%%%%%%%%%%%%%
%%%%%%%%%%%%%%%%%%%%%%%%%%%%

A straightforward perturbative calculation based on the Schwinger-Keldysh formalism \cite{Schwinger,Schwinger2,Keldysh,haug,tatara,kita} gives the statistical average of those magnonic currents $ \langle  {\hat{{\cal{I}}}}_{{\rm{m}}(Q)} \rangle  \equiv  {\cal{I}}_{{\rm{m}}(Q)} $ and the zero-frequency noise  ${\cal{S}}_{{\rm{m}}(Q)}$ up to ${\cal{O}} (J_{\rm{ex}}^2)$ \cite{SeeSuppl};
\begin{subequations}
\begin{eqnarray}
{\cal{I}}_{\rm{m}} &=&  4 (J_{\rm{ex}}S)^2   \sum_{{\mathbf{k}}, k_x^{\prime}}  \int \frac{d\omega }{2 \pi}  
{\rm{Im}} G_{{\rm{L}}, {\mathbf{k}}, \omega }^{\rm{r}}   {\rm{Im}} G_{{\rm{R}}, {\mathbf{k}}^{\prime}, \omega }^{\rm{r}}    \nonumber    \\
&\times &   g \mu _{\rm{B}}  \cdot     \Delta n(\omega ),
 \label{eqn:Im}  \\
%%%%%%%%%%%%%%%%%%%%%%%%%%%%%%%%%%
{\cal{S}}_{\rm{m}} &=&  4 (J_{\rm{ex}}S)^2   \sum_{{\mathbf{k}}, k_x^{\prime}}  \int \frac{d\omega }{2 \pi}  
{\rm{Im}} G_{{\rm{L}}, {\mathbf{k}}, \omega }^{\rm{r}}   {\rm{Im}} G_{{\rm{R}}, {\mathbf{k}}^{\prime}, \omega }^{\rm{r}}    \nonumber    \\
&\times &   (g \mu _{\rm{B}})^2    [2{\cal{F}}_0 (\omega )   +    {\rm{coth}}(\beta \hbar \omega /2) \cdot  \Delta n(\omega )],
 \label{eqn:Sm}  \\
%%%%%%%%%%%%%%%%%%%%%%%%%%%%%%%%%%
{\cal{I}}_{Q} &=&  4 (J_{\rm{ex}}S)^2   \sum_{{\mathbf{k}}, k_x^{\prime}}  \int \frac{d\omega }{2 \pi}  
{\rm{Im}} G_{{\rm{L}}, {\mathbf{k}}, \omega }^{\rm{r}}  {\rm{Im}} G_{{\rm{R}}, {\mathbf{k}}^{\prime}, \omega }^{\rm{r}}    \nonumber    \\
&\times &   \hbar \omega     \cdot    \Delta n(\omega ),
 \label{eqn:IQ}  \\
%%%%%%%%%%%%%%%%%%%%%%%%%%%%%%%%%%
{\cal{S}}_{Q} &=&  4 (J_{\rm{ex}}S)^2   \sum_{{\mathbf{k}}, k_x^{\prime}}  \int \frac{d\omega }{2 \pi}  
{\rm{Im}} G_{{\rm{L}}, {\mathbf{k}}, \omega }^{\rm{r}}   {\rm{Im}} G_{{\rm{R}}, {\mathbf{k}}^{\prime}, \omega }^{\rm{r}}    \nonumber    \\
&\times &   (\hbar \omega )^2   [2{\cal{F}}_0 (\omega )   +    {\rm{coth}}(\beta \hbar \omega /2) \cdot  \Delta n(\omega )],
 \label{eqn:SQ}
%%%%%%%%%%%%%%%%%%%%%%%%%%%%%%%%%%magnonWF,KJD,KSJD,ReviewMagnon
\end{eqnarray}
\end{subequations}
where 
 $   \beta \equiv (k_{\rm{B}}T)^{-1}$, the retarded Green's function $G^{\rm{r}}$,
$  {\cal{F}}_l (\omega ) \equiv  (\hbar \omega )^l {\rm{e}}^{\beta \hbar \omega }/({\rm{e}}^{\beta \hbar \omega }-1)^2  $ for $ l \in  {\mathbb{Z}}$,
and the difference of the Bose-distribution functions between the left FI and the right one
$\Delta n(\omega ) \equiv   n_{\rm{R}}(\omega ) - n_{\rm{L}}(\omega ) $ for $ n(\omega )=({\rm{e}}^{\beta \hbar \omega }-1)^{-1}  $.
%which includes phenomenologically a lifetime of magnons mainly due to nonmagnetic impurity scatterings, being temperature-independent at low temperatures, and thereby we assume that the lifetime of magnons in each FI is identical.

%%%%%%%%%%%%%%%%%
%\section{Onsager coefficients}
%\label{sec:Onsager}
%%%%%%%%%%%%%%%%%
{\it{Onsager coefficients.$-$}}
%%%%%%%%%%%%%%%%%
Assuming the Zeeman term $ g \mu_{\rm{B}} B_{\rm{L(R)}}$ in the presence of an applied strong magnetic field $B_{\rm{L(R)}}$, we expand Eqs. (\ref{eqn:Im})-(\ref{eqn:SQ}) in powers of  $\Delta B  \equiv   B_{\rm{R}} - B_{\rm{L}} $ \cite{magnon2,Trauzettel,Haldane2} 
%\footnote{The resulting magnetization difference from the applied temperature difference works as an effective magnetic field difference \cite{SilsbeeMagnetization,Basso,Basso2,Basso3} which can be identified with the difference of a nonequilibrium magnonic spin chemical potential \cite{WeesNatPhys,MagnonChemicalWees,YacobyChemical}.} 
and $\Delta T   \equiv   T_{\rm{R}} - T_{\rm{L}}$ for 
$\mid \Delta B    \mid  \ll  B_{\rm{L}} \equiv  B$ and $ \mid  \Delta T  \mid  \ll   T_{\rm{L}} \equiv T $.
Within the linear response regime, those magnonic currents and the noise are characterized by the Onsager coefficients, 
$L_{ij}$, ${\cal{S}}_{ij}$ ($i, j = 1, 2$), ${\cal{S}}_{\rm{m}}^{0} $, and ${\cal{S}}_Q^{0 }$ as a function of $B$ and $T$;
\begin{eqnarray}
%%%%%%%%%%%%%%%%%%%%%%%%%%%%%
\begin{pmatrix}
{\cal{I}}_{\rm{m}}  \\    
{\cal{S}}_{\rm{m}}    \\   
{\cal{I}}_{Q}   \\   
{\cal{S}}_Q
\end{pmatrix}
%%%%%%%%%%%%%%%%%%%%%%%%%%%%%
=
%%%%%%%%%%%%%%%%%%%%%%%%%%%%%
\begin{pmatrix}
0  & L_{11} & L_{12}  \\ 
{\cal{S}}_{\rm{m}}^{0}   & {\cal{S}}_{11} & {\cal{S}}_{12}  \\
0  & L_{21} & L_{22}  \\
{\cal{S}}_Q^{0 }   & {\cal{S}}_{21} & {\cal{S}}_{22}  \\
\end{pmatrix}
%%%%%%%%%%%%%%%%%%%%%%%%%%%%%
%%%%%%%%%%%%%%%%%%%%%%%%%%%%%
\begin{pmatrix}
1  \\
- \Delta B  \\  
\Delta T
\end{pmatrix},
%%%%%%%%%%%%%%%%%%%%%%%%%%%%%
\label{eqn:Matrix}
\end{eqnarray}
where (see supplemental material for $ {{\cal{S}}_{11}}$ and $ {{\cal{S}}_{21}}$)
\begin{subequations}
\begin{eqnarray}
L_{11} &=&  \beta (g \mu _{\rm{B}})^2 \int \frac{d\omega }{2 \pi}   {\cal{G}}   {\cal{F}}_0,     \   
L_{22} =   \frac{\beta}{T} \int \frac{d\omega }{2 \pi}   {\cal{G}}   {\cal{F}}_2,
 \label{eqn:L11}  \\
%%%%%%%%%%%%%%%%%%%%%%%%%%%%%%%%%%
L_{12} &=&  \frac{L_{21}}{T} =   \frac{\beta g \mu _{\rm{B}}}{T} \int \frac{d\omega }{2 \pi}   {\cal{G}}   {\cal{F}}_1 ,
 \label{eqn:L12} \\
%%%%%%%%%%%%%%%%%%%%%%%%%%%%%%%%%%
%L_{22} &=&   \frac{\beta}{T} \int \frac{d\omega }{2 \pi}   {\cal{G}}   {\cal{F}}_2,
 %\label{eqn:L22} \\
%%%%%%%%%%%%%%%%%%%%%%%%%%%%%%%%%%
{\cal{S}}_{12} &=& \frac{\beta (g \mu _{\rm{B}})^2}{T} \int \frac{d\omega }{2 \pi}   {\cal{G}}   {\cal{F}}_1  {\rm{coth}}(\beta \hbar \omega /2),
 \label{eqn:S12} \\
%%%%%%%%%%%%%%%%%%%%%%%%%%%%%%%%%%
{\cal{S}}_{22} &=& \frac{\beta}{T} \int \frac{d\omega }{2 \pi}   {\cal{G}}   {\cal{F}}_3  {\rm{coth}}(\beta \hbar \omega /2),
 \label{eqn:S22} \\ 
 %%%%%%%%%%%%%%%%%%%%%%%%%%%%%%%%%%
{\cal{S}}_{\rm{m}}^{0} &=& 2 (g \mu _{\rm{B}})^2 \int \frac{d\omega }{2 \pi}   {\cal{G}}  {\cal{F}}_0,   \   \  
{\cal{S}}_{Q}^{0 } = 2 \int \frac{d\omega }{2 \pi}   {\cal{G}}   {\cal{F}}_2,
 \label{eqn:Smstar}
%%%%%%%%%%%%%%%%%%%%%%%%%%%%%%%%%%
%{\cal{S}}_{Q}^{\ast } &=& 2 \int \frac{d\omega }{2 \pi}   {\cal{G}}   {\cal{F}}_2,
 %\label{eqn:SQstar} 
%%%%%%%%%%%%%%%%%%%%%%%%%%%%%%%%%%
\end{eqnarray}
\end{subequations}
 %$  {\cal{F}}_l (\omega ) \equiv  (\hbar \omega )^l {\rm{e}}^{\beta \hbar \omega }/({\rm{e}}^{\beta \hbar \omega }-1)^2  $ for $ l \in  {\mathbb{Z}}$,
 and ${\cal{G}} = {\cal{G}}(B)  \equiv  4 (J_{\rm{ex}}S)^2   \sum_{{\mathbf{k}}, k_x^{\prime}} {\rm{Im}} G_{{\rm{L}}, {\mathbf{k}}, \omega }^{\rm{r}}  {\rm{Im}} G_{{\rm{R}}, {\mathbf{k}}^{\prime}, \omega }^{\rm{r}}\mid _{B_{\rm{R}}=B}$.
Note that throughout this paper we focus on low temperatures in the presence of the applied strong magnetic field $0 \not= k_{\rm{B}} T \ll   g \mu_{\rm{B}} B $ where effects of magnon-magnon (-phonon) interactions and the resulting effective magnetic field become negligibly small \cite{Tmagnonphonon,adachiphonon}, and thereby we assume phenomenologically a temperature-independent lifetime of magnons in $ G^{\rm{r}}$ (i.e., $\partial  {\cal{G}}/\partial T =0 $) \footnote{
Even if the lifetime is temperature-dependent, the results qualitatively remain valid in the magnonic shot noise regime \cite{SeeSuppl}.
}
mainly due to nonmagnetic impurity scatterings.
%%%%%%%%%
From Eq. (\ref{eqn:L12}) the coefficients $L_{ij}$ for the currents are seen to satisfy the Onsager relations $L_{21} = T L_{12} $.
%%%%%%%%%
We remark that in contrast to the current, the noise consists of two parts [Eq. (\ref{eqn:Matrix})]; the equilibrium term ${\cal{S}}_{{\rm{m}}(Q)}^{0 } $ and the nonequilibrium component ${\cal{S}}_{ij}$.
Since the noise is the second moment of the current operator, it does not vanish even when no current flows 
$\langle  {\hat{{\cal{I}}}}_{{\rm{m}}(Q)} \rangle =0$, 
i.e.  $ \Delta B = \Delta T =0  $.
This property intrinsic to the noise is described by the equilibrium term ${\cal{S}}_{{\rm{m}}(Q)}^{0} $, namely, the thermal noise (or the Johnson-Nyquist noise) \cite{ThermalNoise,ThermalNoise2}, which arises from the ${\cal{F}}_0 $-term in Eqs. (\ref{eqn:Sm}) and (\ref{eqn:SQ}).

%%%%%%%%%%%%%%%%%
%\section{Thermomagnetic relations}
%\label{sec:relation}
%%%%%%%%%%%%%%%%%%%%%
{\it{Thermomagnetic relations.$-$}}
%%%%%%%%%%%%%%%%%%%%%
From Eqs. (\ref{eqn:L11})-(\ref{eqn:Smstar}) we obtain thermomagnetic relations between the magnonic currents and the noise.
The ratios of the thermal noise to the Onsager coefficients $L_{ij}$ for the magnonic spin and heat currents,
$ {{\cal{S}}_{\rm{m}}^{0 }}/{L_{11}}$ and $ {{\cal{S}}_{Q}^{0 }}/{L_{22}}$, satisfy the Onsager relations;
\begin{eqnarray}
\frac{{\cal{S}}_{\rm{m}}^{0 }}{L_{11}} = \frac{1}{T}  \frac{{\cal{S}}_{Q}^{0 }}{L_{22}}= 2 k_{\rm{B}}T.
\label{eqn:OnsagerS} 
\end{eqnarray}
The ratios exhibit a {\it universal} behavior, i.e. they are completely independent of any material parameters (e.g. $g$-factor, spin length, and exchange interaction, etc.) and are solely determined by temperature.
%%%%%%%%%%%%%
In the low-temperature limit $ 0 \not= k_{\rm{B}} T \ll   g \mu_{\rm{B}} B $, due to the Zeeman energy $   g \mu_{\rm{B}}B $  the ratio of the coefficient for the thermally-induced nonequilibrium spin noise ${{\cal{S}}_{12}}$ 
to that for the spin current ${L_{12}}$ reduces to
\begin{eqnarray}
\frac{{\cal{S}}_{12}}{L_{12}} =    g \mu_{\rm{B}}  \frac{\int \frac{d\omega }{2 \pi}   {\cal{G}}   {\cal{F}}_1  {\rm{coth}}(\beta \hbar \omega /2)}{\int \frac{d\omega }{2 \pi}   {\cal{G}}  {\cal{F}}_1}  
\stackrel{\rightarrow }{=}  g \mu_{\rm{B}},
 \label{eqn:FanoMagnon} 
\end{eqnarray}
and becomes independent of any material parameters except the $g$-factor being material specific.
%%%%%%%%%%%%%%%%%
Thus, the Onsager coefficients $L_{ij}$ for the magnonic currents are described by the noise language;
\begin{eqnarray}
L_{11} = \frac{{\cal{S}}_{\rm{m}}^{0 }}{2 k_{\rm{B}}T},  \  \     
L_{12} = \frac{L_{21}}{T}   \stackrel{\rightarrow }{=}  \frac{{\cal{S}}_{12}}{g \mu_{\rm{B}}}, \  \      
L_{22} = \frac{{\cal{S}}_{Q}^{0 }}{2 k_{\rm{B}}T^2}.
 \label{eqn:LSconnect} 
\end{eqnarray}
Note that being independent of microscopic details of the FIs (e.g. the energy dispersion of magnons), 
those thermomagnetic relations [Eq. (\ref{eqn:LSconnect})] between the magnonic currents and the noise hold.

%%%%%%%%%%%%%%%%%
%\section{Noise and WF law}
%\label{sec:WF}
%%%%%%%%%%%%%%%%%
{\it{Noise and WF law.$-$}}
%%%%%%%%%%%%%%%%%
In analogy to charge transport in metals \cite{LandauWF,AMermin}, the coefficient $L_{11}$ is identified with the magnonic spin conductance  
$G \equiv  L_{11}$ \cite{MagnonG}
and the thermal conductance $K$ is defined by $K \equiv  L_{22}-L_{12}L_{21}/L_{11} $ \cite{magnonWF,KJD,KSJD,ReviewMagnon}.
%Since magnons are bosonic quasiparticles with a magnetic dipole moment $g \mu_{\rm{B}} {\mathbf{e}}_z$, the off-diagonal contributions $ L_{12}L_{21}/L_{11} $ in the thermal conductance are not suppressed. This is in contrast to electrons (i.e. fermions) in metals \cite{LandauWF,AMermin}.
%%%%%%%%%%%
In the low-temperature limit $ 0 \not= k_{\rm{B}} T \ll   g \mu_{\rm{B}} B $, the magnonic WF law for the junction holds \cite{magnonWF};
${K}/{G}=  (L_{22}-L_{12}L_{21}/L_{11})/L_{11}   \stackrel{\rightarrow }{=}   (k_{\rm{B}}/g \mu _{\rm{B}})^2 T$.
From the WF law and Eq. (\ref{eqn:LSconnect}) we obtain 
\begin{eqnarray}
\frac{{\cal{S}}_Q^{0 }}{{\cal{S}}_{\rm{m}}^{0 }} - \Big(\frac{2  k_{\rm{B}}T^2}{g \mu _{\rm{B}}} \Big)^2     
\Big(\frac{{\cal{S}}_{12}}{{\cal{S}}_{\rm{m}}^{0 }}\Big)^2 \stackrel{\rightarrow }{=}   \Big(\frac{k_{\rm{B}}T}{g \mu _{\rm{B}}}\Big)^2. 
 \label{eqn:WFnoise} 
\end{eqnarray}
Thus the magnonic WF law for the ferromagnetic insulating junction, characterizing the universal thermomagnetic properties of the magnonic spin and heat currents, 
is explained by the noise language.
Eq. (\ref{eqn:WFnoise}) is seen to describe the universal relations between the thermal noise ${\cal{S}}_{{\rm{m}}(Q)}^{0 }$ and the nonequilibrium spin noise ${\cal{S}}_{12}$, and 
we refer to Eq. (\ref{eqn:WFnoise}) as the {\it{WF law for the magnonic thermal noise}}.
%%%%%%%%%%%%%
From Eqs. (\ref{eqn:WFnoise}) and (\ref{eqn:FanoMagnon}) we obtain
$L_{12}= {L_{21}}/{T}  \stackrel{\rightarrow }{=}  ({{\cal{S}}_{\rm{m}}^{0 }}/{2  k_{\rm{B}}T^2})
\sqrt{{{\cal{S}}_Q^{0 }}/{{\cal{S}}_{\rm{m}}^{0 }} - ({k_{\rm{B}}T}/{g \mu _{\rm{B}}})^2}$,
%\begin{eqnarray}
%L_{12}= \frac{L_{21}}{T}  \stackrel{\rightarrow }{=}  \frac{{\cal{S}}_{\rm{m}}^{\ast }}{2  k_{\rm{B}}T^2}
%\sqrt{\frac{{\cal{S}}_Q^{\ast }}{{\cal{S}}_{\rm{m}}^{\ast }} - \Big(\frac{k_{\rm{B}}T}{g \mu _{\rm{B}}}\Big)^2},
%\label{eqn:WFnoiseL12} 
%\end{eqnarray}
and see that  the coefficients $L_{12}$ and $L_{21}$ are characterized by the thermal noise 
${\cal{S}}_{{\rm{m}}(Q)}^{0 } $ (i.e. without using the nonequilibrium term $ {\cal{S}}_{12}$).
Thus making use of the magnonic WF law, all the Onsager coefficients $L_{ij}$ for the magnonic spin and heat currents 
[Eq. (\ref{eqn:LSconnect})] are described solely in terms of the thermal noise $L_{ij} = L_{ij}({\cal{S}}_{\rm{m}}^{0 }, {\cal{S}}_Q^{0 })$.

%%%%%%%%%%%%%%%%%
%\section{Noise and Seebeck coefficient}
%\label{sec:Seebeck}
%%%%%%%%%%%%%%%%%%%%%%
{\it{Noise and Seebeck coefficient.$-$}}
%%%%%%%%%%%%%%%%%%%%%%
Since the magnonic Seebeck coefficient \cite{magnonWF} reduces to $  L_{12}/L_{11} \stackrel{\rightarrow }{=}  B/T$ in the low-temperature limit $0 \not=  k_{\rm{B}} T \ll   g \mu_{\rm{B}} B $, Eq. (\ref{eqn:LSconnect}) provides
\begin{eqnarray}
\frac{{\cal{S}}_{12}}{{\cal{S}}_{\rm{m}}^{0 }}  
\stackrel{\rightarrow }{=}  \frac{g \mu_{\rm{B}}}{2 k_{\rm{B}}T}\frac{L_{12}}{L_{11}}
\stackrel{\rightarrow }{=}  \frac{g \mu_{\rm{B}}B}{2 k_{\rm{B}}T^2},
 \label{eqn:WFnoise2} 
\end{eqnarray}
where the ratio becomes independent of any material parameters except the $g$-factor being material specific.
From Eqs. (\ref{eqn:WFnoise2}) and (\ref{eqn:WFnoise}) we obtain
\begin{eqnarray}
\frac{{\cal{S}}_Q^{0 }}{{\cal{S}}_{\rm{m}}^{0 }} 
 \stackrel{\rightarrow }{=}  B^2 \Big[1+\Big(\frac{ k_{\rm{B}}T}{g \mu_{\rm{B}}B}  \Big)^2 \Big]
 \stackrel{\rightarrow }{=}  B^2.
\label{eqn:WFnoise3} 
\end{eqnarray}
Thus the ratio between each thermal noise $ {{\cal{S}}_Q^{0 }}/{{\cal{S}}_{\rm{m}}^{0 }} $ exhibits a universal behavior at low temperatures, i.e. it is completely independent of any material parameters (e.g. $g$-factor, spin length, and exchange interaction, etc.) and is solely determined by the applied magnetic field.
%%%%%%%%%%%%%%%%
From Eqs. (\ref{eqn:WFnoise2}) and (\ref{eqn:WFnoise3}),
each thermal noise ${\cal{S}}_{{\rm{m}}(Q)}^{0 }$ is seen to be characterized by the nonequilibrium spin noise ${\cal{S}}_{12}$;
${{\cal{S}}_{\rm{m}}^{0 }}  \stackrel{\rightarrow }{=}  ({2 k_{\rm{B}}T^2}/{g \mu_{\rm{B}}B}){{\cal{S}}_{12}}$,
$ {{\cal{S}}_Q^{0 }}    \stackrel{\rightarrow }{=} [1+({ k_{\rm{B}}T}/{g \mu_{\rm{B}}B})^2]  ({2 k_{\rm{B}}T^2 B}/{g \mu_{\rm{B}}}){{\cal{S}}_{12}}
\stackrel{\rightarrow }{=}  ({2 k_{\rm{B}}T^2 B}/{g \mu_{\rm{B}}}){{\cal{S}}_{12}}  $.
%\begin{eqnarray}
%{{\cal{S}}_{\rm{m}}^{\ast }}  \stackrel{\rightarrow }{=}  \frac{2 k_{\rm{B}}T^2}{g \mu_{\rm{B}}B}{{\cal{S}}_{12}},   \   \    \   
%{{\cal{S}}_Q^{\ast }}  \stackrel{\rightarrow }{=}  \frac{2 k_{\rm{B}}T^2 B}{g \mu_{\rm{B}}}{{\cal{S}}_{12}}.
%\label{eqn:WFnoise4} 
%\end{eqnarray}
Thus making use of both the magnonic WF law and the Seebeck coefficient [Eqs. (\ref{eqn:WFnoise}) and (\ref{eqn:WFnoise2})], all the Onsager coefficients $L_{ij}$ for the magnonic spin and heat currents can be described solely by each noise  [Eq. (\ref{eqn:LSconnect})];
$L_{ij} = L_{ij}({\cal{S}}_{\rm{m}}^{0 })= L_{ij}({\cal{S}}_Q^{0 })= L_{ij}({\cal{S}}_{12})$.

%Note that Eqs. (\ref{eqn:WFnoise3}) and (\ref{eqn:WFnoiseL12}) give 
%$L_{12}= L_{21}/T  \stackrel{\rightarrow }{=}  (B/2  k_{\rm{B}}T^2){\cal{S}}_{\rm{m}}^{\ast }$, and the coefficients $L_{12}$ and $L_{21}$ become characterized solely by ${\cal{S}}_{\rm{m}}^{\ast } $ at low temperatures.

%%%%%%%%%%%%%%%%%
%\section{Noise and thermal conductance}
%\label{sec:NoiseK}
%%%%%%%%%%%%%%%%%
{\it{Noise and thermal conductance.$-$}}
%%%%%%%%%%%%%%%%%
Making use of those thermomagnetic relations for the magnonic currents (i.e. the WF law, the Seebeck coefficient, and the Onsager relations) \cite{magnonWF}, 
each coefficient $L_{ij}$ is rewritten solely in terms of the thermal conductance $K$ at low temperatures $L_{ij} \stackrel{\rightarrow }{=} L_{ij}(K) $;
%%%%%%%%%%%%%%%%%%%%%%%%%%%%%%%%%%%%%%%%%%%%%%
$  L_{11} \stackrel{\rightarrow }{=} ({1}/{T}) ({g \mu _{\rm{B}}}/{k_{\rm{B}}})^2  K  $,
$  L_{12} =  {L_{21}}/{T}  \stackrel{\rightarrow }{=}  ({B}/{T^2}) ({g \mu _{\rm{B}}}/{k_{\rm{B}}})^2  K   $,
$  L_{22} \stackrel{\rightarrow }{=}  [1+({g \mu _{\rm{B}} B}/{k_{\rm{B}}T})^2]  K    
\stackrel{\rightarrow }{=} ({g \mu _{\rm{B}} B}/{k_{\rm{B}}T})^2  K  $.
%%%%%%%%%%%%%%%%%%%%%%%%%%%%%%%%%%%%%%%%%%%%%
From those we obtain
\begin{eqnarray}
\frac{L_{22}}{L_{11}}  \stackrel{\rightarrow }{=}   \Big[1+\Big(\frac{ k_{\rm{B}}T}{g \mu_{\rm{B}}B} \Big)^2 \Big]   \frac{B^2}{T}
  \stackrel{\rightarrow }{=} \frac{B^2}{T} \propto  \frac{1}{T},
\end{eqnarray}
and see that in contrast to the fermionic counterpart (i.e. the WF law for charge transport in metals \cite{WFgermany}), the ratio does not exhibit a $T$-linear behavior.
This illustrates the significance of the off-diagonal contributions 
\footnote{This is in contrast to the fermionic counterpart (i.e. the WF law for charge transport in metals \cite{WFgermany}), where the off-diagonal contributions are strongly suppressed by the sharp Fermi surface of fermions even at room temperatures being much smaller than the Fermi energy \cite{LandauWF,AMermin}, and thereby those off-diagonal contributions become negligibly small.}
$ L_{12}L_{21}/L_{11} $ in the thermal conductance $K  \equiv  L_{22}-L_{12}L_{21}/L_{11}$
to the magnonic WF law ${K}/{G}  \stackrel{\rightarrow }{=}   (k_{\rm{B}}/g \mu _{\rm{B}})^2 T$.
Those off-diagonal elements account for the counter-current due to the resulting magnetization difference from the applied temperature difference \footnote{At low temperatures, while being much smaller than the external strong magnetic field, the resulting magnetization difference from the applied temperature difference plays a role of an effective magnetic field difference \cite{SilsbeeMagnetization,Basso,Basso2,Basso3} which can be identified with the difference of a nonequilibrium magnonic spin chemical potential \cite{WeesNatPhys,MagnonChemicalWees,YacobyChemical,MagnonG}.},
and by taking the contribution into account appropriately \cite{magnonWF,KJD,KSJD,ReviewMagnon}, the linear-in-$T$ behavior of the WF law holds in the same way for magnons despite the fundamental difference of the quantum-statistical properties between bosons and fermions.
%%%%%%%%%%%%%%
Finally, from Eq. (\ref{eqn:LSconnect}) and $L_{ij}(K) $  we obtain
\begin{eqnarray}
{\cal{S}}_{\rm{m}}^{0 } \stackrel{\rightarrow }{=}   \frac{{\cal{S}}_{Q}^{0 }}{B^2 \big[1+\big(\frac{ k_{\rm{B}}T}{g \mu_{\rm{B}}B}  \big)^2 \big]}    
\stackrel{\rightarrow }{=}   2 \frac{(g \mu _{\rm{B}})^2}{k_{\rm{B}}}  K,    
%%%%%%%%%%%%%%%%%%%%
%{\cal{S}}_{Q}^{\ast } &\stackrel{\rightarrow }{=}&    2 \frac{(g \mu _{\rm{B}}B)^2}{k_{\rm{B}}}  K,
%\label{eqn:SQK}     \\
%%%%%%%%%%%%%%%%%%%%
{\cal{S}}_{12}  \stackrel{\rightarrow }{=}  \frac{(g \mu _{\rm{B}})^3 B}{(k_{\rm{B}}T)^2}  K.    \     
\label{eqn:S12K}    
\end{eqnarray}
Thus each noise is described in terms of the thermal conductance $K$ being measurable with current device and measurement techniques.

%%%%%%%%%%%%%%%%%%%%%%%%
%\section{Magnonic shot noise}
%\label{sec:shot}
%%%%%%%%%%%%%%%%%%%%%%%%
{\it{Magnonic shot noise.$-$}}
%%%%%%%%%%%%%%%%%%%%%%%%
When $  2{\cal{F}}_0 (\omega )   \ll     {\rm{coth}}(\beta \hbar \omega /2) \mid   \Delta n(\omega )\mid   $ in Eqs. (\ref{eqn:Sm}) and (\ref{eqn:SQ}), 
the thermal noise $ {\cal{S}}_{{\rm{m}}(Q)}^{0 }  $ is seen to be negligibly small and the nonequilibrium terms ${\cal{S}}_{ij}$ become the leading components of the noise $ {\cal{S}}_{{\rm{m}}(Q)}  $ [Eq. (\ref{eqn:Matrix})].
Thereby in analogy to the shot noise for electron transport \cite{ShotNoise1918,NoiseRev}, we refer to the nonequilibrium part described by ${\cal{S}}_{ij}$ as the {\it{magnonic shot noise}}.
%%%%%%%%%%%%%%
In the low-temperature limit (i.e. $0 \not= k_{\rm{B}} T \ll   g \mu_{\rm{B}} B $) of the magnonic shot noise regime, the nonequilibrium components ${\cal{S}}_{11}$ and ${\cal{S}}_{21}$ become \cite{SeeSuppl}
\begin{eqnarray}
\frac{{\cal{S}}_{11}}{L_{11}}  \stackrel{\rightarrow }{=}    \frac{{\cal{S}}_{21}}{T L_{22}} \stackrel{\rightarrow }{=}  g \mu _{\rm{B}}, 
\label{eqn:S11S21}    
\end{eqnarray}
and the ratios reduce to the value being independent of any material parameters except the $g$-factor which is material specific.
Using Eqs. (\ref{eqn:S11S21}) and (\ref{eqn:FanoMagnon}), the magnonic WF law for the currents $L_{ij}$ is explained solely by the nonequilibrium components $  {\cal{S}}_{ij} $ in the magnonic shot noise regime;
\begin{eqnarray}
 \frac{{\cal{S}}_{21}}{{\cal{S}}_{11}}  - T^2 \Big(\frac{{\cal{S}}_{12}}{{\cal{S}}_{11}}\Big)^2  \stackrel{\rightarrow }{=}  
\Big(\frac{k_{\rm{B}}T}{g \mu _{\rm{B}}}\Big)^2. 
\label{eqn:WFnonequilibriumNoise}    
\end{eqnarray}
This describes the universal relations among the nonequilibrium spin and heat noise ${\cal{S}}_{ij}$, and 
we refer to Eq. (\ref{eqn:WFnonequilibriumNoise}) as the {\it{WF law for the magnonic shot noise}}.
%%%%%%%%%%%%%%%%%%%%%%%%%%%%%%%%%%%%%%%%%
From Eqs. (\ref{eqn:S11S21}) and (\ref{eqn:LSconnect}) we obtain the relations between the nonequilibrium noise ${\cal{S}}_{11(21)}$ and the thermal noise $  {\cal{S}}_{{\rm{m}}(Q)}^{0 }  $;
\begin{eqnarray}
 \frac{{\cal{S}}_{11}}{{\cal{S}}_{\rm{m}}^{0}}   \stackrel{\rightarrow }{=} \frac{{\cal{S}}_{21}}{{\cal{S}}_{Q}^{0 }}      
\stackrel{\rightarrow }{=}    \frac{g \mu _{\rm{B}}}{2 k_{\rm{B}}T}.
\label{eqn:S11SmS21SQ}    
\end{eqnarray}
Finally, from Eq. (\ref{eqn:S12K}), each nonequilibrium noise ${\cal{S}}_{11(21)}$ is described by the thermal conductance $K$;
\begin{eqnarray}
 {{\cal{S}}_{11}}   \stackrel{\rightarrow }{=}  \frac{(g \mu _{\rm{B}})^3}{k_{\rm{B}}^2 T}K,   \    \    
  {{\cal{S}}_{21}}   \stackrel{\rightarrow }{=}  g \mu _{\rm{B}}T \Big[1+\Big(\frac{g \mu_{\rm{B}}B}{ k_{\rm{B}}T} \Big)^2 \Big] K.
\label{eqn:S11S21K}    
\end{eqnarray}
Note that the WF law for the magnonic shot noise [Eq. (\ref{eqn:WFnonequilibriumNoise}) with Eq. (\ref{eqn:S11SmS21SQ})] reproduces 
the one for the thermal noise $  {\cal{S}}_{{\rm{m}}(Q)}^{0 }  $ [Eq. (\ref{eqn:WFnoise})].

%%%%%%%%%%%%%%%%%%%%%%%%
%\section{Magnonic spin-Fano factor}
%\label{sec:Fano}
%%%%%%%%%%%%%%%%%%%%%%%%
{\it{Magnonic spin-Fano factor.$-$}}
%%%%%%%%%%%%%%%%%%%%%%%%
So far we have considered the magnonic currents and the noise within the linear response regime in respect to $ \Delta B$ and $ \Delta T$ [Eq. (\ref{eqn:Matrix})].
Focusing on a larger temperature difference $ \Delta T$
(i.e. $ g \mu_{\rm{B}} \mid  \Delta  B \mid  \ll     k_{\rm{B}} \mid  \Delta T  \mid  \ll  k_{\rm{B}}T $), 
the next leading contribution is assumed to be the response to $(\Delta T)^2 $, i.e. the nonlinear response in respect to $ \Delta T$.
Expanding Eqs. (\ref{eqn:Im})-(\ref{eqn:SQ}) again in powers of $  \Delta T  $ but up to ${\cal{O}}\big((\Delta T)^2\big) $,
the Onsager matrix Eq. (\ref{eqn:Matrix}) is extended into
\begin{eqnarray}
%%%%%%%%%%%%%%%%%%%%%%%%%%%%%
\begin{pmatrix}
{\cal{I}}_{\rm{m}}  \\    
{\cal{S}}_{\rm{m}}    \\   
{\cal{I}}_{Q}   \\   
{\cal{S}}_Q
\end{pmatrix}
%%%%%%%%%%%%%%%%%%%%%%%%%%%%%
=
%%%%%%%%%%%%%%%%%%%%%%%%%%%%%
\begin{pmatrix}
0  & L_{11} & L_{12} & L_{13} \\ 
{\cal{S}}_{\rm{m}}^{0 }   & {\cal{S}}_{11} & {\cal{S}}_{12} & {\cal{S}}_{13}  \\
0  & L_{21} & L_{22} & L_{23}  \\
{\cal{S}}_Q^{0 }   & {\cal{S}}_{21} & {\cal{S}}_{22} & {\cal{S}}_{23}  \\
\end{pmatrix}
%%%%%%%%%%%%%%%%%%%%%%%%%%%%%
%%%%%%%%%%%%%%%%%%%%%%%%%%%%%
\begin{pmatrix}
1  \\
- \Delta B  \\  
\Delta T \\
(\Delta T)^2
\end{pmatrix},
%%%%%%%%%%%%%%%%%%%%%%%%%%%%%
\label{eqn:Matrix2}
\end{eqnarray}
where \cite{SeeSuppl}
$    L_{13} =   - ({\beta g \mu _{\rm{B}}}/{T^2})   \int ({d\omega }/{2 \pi})   {\cal{G}} {\cal{F}}_1    
            +   ({\beta^2 g \mu _{\rm{B}}}/{2 T^2})  \int ({d\omega }/{2 \pi})   {\cal{G}}  {\cal{F}}_2  {\rm{coth}}(\beta \hbar \omega /2)   $, and
$ {{\cal{S}}_{13}} =  - [{\beta (g \mu _{\rm{B}})^2}/{T^2}]   \int ({d\omega }/{2 \pi})   {\cal{G}} {\cal{F}}_1  {\rm{coth}}(\beta \hbar \omega /2) 
              +   [{(\beta g \mu _{\rm{B}})^2}/{2 T^2}]  \int ({d\omega }/{2 \pi})   {\cal{G}}      {\cal{F}}_2  {\rm{coth}}^2(\beta \hbar \omega /2) $.
%%%%%%%%%%%%%%%%%%%%%%%%%%%%%%%%%%%%%%%%%%%%%%%%%%%%%%%%%
In analogy to the Fano factor of the noise (i.e. noise-to-current ratio) for electron transport \cite{DLnoise3}, 
we identify the ratio of the spin noise to the spin current with the magnonic spin-Fano factor,
and refer to $ F_{\rm{s}} (\Delta T ) \equiv {\cal{S}}_{\rm{m}}(\Delta B=0)/{\cal{I}}_{\rm{m}}(\Delta B=0)   $ as the thermally-induced magnonic spin-Fano factor.
%%%%%%%%%%%%%%%%%%%%%%%%%%%
In the magnonic shot noise regime, the thermal noise $ {\cal{S}}_{{\rm{m}}}^{0 }  $ is negligibly small and it becomes
\begin{eqnarray}
F_{\rm{s}} (\Delta T ) 
&= & \frac{ {{\cal{S}}_{12}}\Delta T + {{\cal{S}}_{13}} (\Delta T)^2}{L_{12} \Delta T+L_{13} (\Delta T)^2}   \equiv  g   \mu _{\rm{B}}\mid _{\rm{eff}},
\label{eqn:spinFanoNonlinear}      
\end{eqnarray}
where the effective $g$-factor  and the Bohr magneton are introduced and defined as the product $ g   \mu _{\rm{B}}\mid _{\rm{eff}} $ being
$  g   \mu _{\rm{B}}\mid _{\rm{eff}}  \not= g  \mu _{\rm{B}} $ in general.
%%%%%%%%%%%%%%%%%%%%%%%%
However, in the low-temperature limit $0 \not= k_{\rm{B}} T \ll   g \mu_{\rm{B}} B $, the ratio of the nonlinear response 
${{{\cal{S}}_{13}}}/{L_{13}}$ reduces to the same value $g \mu _{\rm{B}}$ with the one in the linear response regime;
$ {{{\cal{S}}_{13}}}/{L_{13}}  \stackrel{\rightarrow }{=}  {{{\cal{S}}_{12}}}/{L_{12}}   \stackrel{\rightarrow }{=}  g \mu _{\rm{B}}    $. 
%%%%%%%%%%%%%%%%%%%%%%%%%%%%%%%%
Thus in the low-temperature limit of the magnonic shot noise regime, 
$ {\Delta T}/{T}   \gg    {2 k_{\rm{B}}T}/{g \mu _{\rm{B}}B} $ (e.g. $T=10$mK, $\Delta T=1$mK, $B=0.5$T),
the magnonic spin-Fano factor exhibits the universal behavior even beyond the linear response regime, 
$ F_{\rm{s}} (\Delta T ) =  g   \mu _{\rm{B}}\mid _{\rm{eff}}  \stackrel{\rightarrow }{=} g  \mu _{\rm{B}} $, 
being independent of any material parameters except the $g$-factor which is material specific.
%%%%%%%%%%%%%%%
Note that from Eqs. (\ref{eqn:Im}) and (\ref{eqn:Sm}) we obtain
$ {\cal{S}}_{\rm{m}}/{\cal{I}}_{\rm{m}} \stackrel{\rightarrow }{=}   g \mu _{\rm{B}}  $ in the regime
and see that the universal behavior of the magnonic spin-Fano factor holds in any orders of $\Delta T$.

%%%%%%%%%%%%%%%%%
%\section{Toward measurement}
%\label{sec:measure}
%%%%%%%%%%%%%%%%%
{\it{Toward measurement.$-$}}
%%%%%%%%%%%%%%%%%
Spin currents and magnetic fluctuations can be measured by using the ISHE \cite{ISHEadd,ishe,ISHEadd2,spinwave,WeesNatPhys} and 
the muon spin rotation and relaxation ($\mu$SR) \cite{muSR,muSR2}, respectively.
Ref. \cite{SpinNoiseMeasure} reported measurements of a spin current noise as the ISHE-induced voltage noise. 
Making use of those measurement techniques and Nitrogen-vacancy (NV) center \cite{YacobyChemical,NVreview}, etc., while being challenging, it will be interesting to test our theoretical predictions experimentally.

%%%%%%%%%%%%%%%%%
%\section{Conclusion}
%\label{sec:conclusion}
%%%%%%%%%%%%%%%%%
{\it{Conclusion.$-$}}
%%%%%%%%%%%%%%%%%
We have established the Onsager relations between the magnonic currents and the zero-frequency noise, and have studied the mutual relations among magnetic momentum, heat, and fluctuations of propagating magnons in the ferromagnetic insulating junction.
%%%%%
Using the Onsager relations, we have derived the Wiedemann-Franz law for the magnonic thermal noise and the one for the magnonic shot noise,
and have discovered the {\it{universal}} behavior of the noise, i.e., being independent of material parameters except the $g$-factor.
%%%%%
Moreover, we showed that the magnonic spin-Fano factor reduces to the universal value in the low-temperature limit and it remains valid even beyond the linear response regime.
%%%%%
We hope our work serves as a bridge between magnon-based spintronics and mesoscopic physics, and provides a new direction in the field of magnonics.

%\begin{acknowledgments}
%\section{Acknowledgments}
{\it{Acknowledgments.$-$}}
We thank H. Chudo, T. Kato, S. Takayoshi, S. Maekawa, S. E. Nigg, and D. Loss for helpful discussions.
This work (KN) is financially supported by Leading Initiative for Excellent Young Researchers, MEXT, Japan.

\appendix

%%%%%%%%%%%%%%
%\begin{widetext}
\begin{center}
\textbf{
Supplemental Material for ``Magnonic Noise and Wiedemann-Franz Law''
}
\end{center}
%\end{widetext}
%%%%%%%%%%%%%%

In this supplemental material, we provide some details on the Onsager coefficients for the magnonic current and the noise.

%%%%%%%%%%%%%%%%%%%%%%%%%%%%%%%%%
 \section{Magnonic current and noise}
\label{sec:Appe}
%%%%%%%%%%%%%%%%%%%%%%%%%%%%%%%%%

Using the Schwinger-Keldysh formalism (i.e., the nonequilibrium Green's function method) \cite{Schwinger,Schwinger2,Keldysh,haug,tatara,kita} 
and treating the tunneling Hamiltonian ${\cal{H}}_{\rm{ex}}$ perturbatively,
the magnonic spin (heat) current ${\cal{I}}_{{\rm{m}}(Q)}$ and the noise ${\cal{S}}_{{\rm{m}}(Q)}$ can be evaluated up to ${\cal{O}} (J_{\rm{ex}}^2)$;
${\cal{I}}_{\rm{m}} =   (J_{\rm{ex}}S)^2   \sum_{{\mathbf{k}}, k_x^{\prime}}  \int  ({d\omega }/{2 \pi})  g \mu _{\rm{B}}
(G_{{\rm{L}}}^{\rm{<}}  G_{{\rm{R}}}^{\rm{>}}   - G_{{\rm{L}}}^{\rm{>}}  G_{{\rm{R}}}^{\rm{<}})$,
%%%%%%%%%%%%%%%%%%%%%%%%%%%%%%%%%%
${\cal{S}}_{\rm{m}} =   - (J_{\rm{ex}}S)^2   \sum_{{\mathbf{k}}, k_x^{\prime}}  \int   ({d\omega }/{2 \pi})  (g \mu _{\rm{B}})^2
(G_{{\rm{L}}}^{\rm{<}}  G_{{\rm{R}}}^{\rm{>}}  + G_{{\rm{L}}}^{\rm{>}}  G_{{\rm{R}}}^{\rm{<}})$,
%%%%%%%%%%%%%%%%%%%%%%%%%%%%%%%%%%
${\cal{I}}_{Q} =   (J_{\rm{ex}}S)^2   \sum_{{\mathbf{k}}, k_x^{\prime}}  \int  ({d\omega }/{2 \pi})  \hbar \omega 
(G_{{\rm{L}}}^{\rm{<}}  G_{{\rm{R}}}^{\rm{>}}   - G_{{\rm{L}}}^{\rm{>}}  G_{{\rm{R}}}^{\rm{<}})$,
%%%%%%%%%%%%%%%%%%%%%%%%%%%%%%%%%%
${\cal{S}}_{Q} =  - (J_{\rm{ex}}S)^2   \sum_{{\mathbf{k}}, k_x^{\prime}}  \int     ({d\omega }/{2 \pi})  (\hbar \omega )^2
(G_{{\rm{L}}}^{\rm{<}}  G_{{\rm{R}}}^{\rm{>}}  + G_{{\rm{L}}}^{\rm{>}}  G_{{\rm{R}}}^{\rm{<}})$,
%%%%%%%%%%%%%%%%%%%%%%%%%%%%%%%%%%
where $ G_{{\rm{L}}}^{<(>)} \equiv    G_{{\rm{L}}, {\mathbf{k}}, \omega }^{<(>)}  $ and 
$ G_{{\rm{R}}}^{<(>)}  \equiv     G_{{\rm{R}}, {\mathbf{k}}^{\prime}, \omega }^{<(>)}  $
are the bosonic lesser (greater) Green's functions of magnons for the left and right FIs, respectively.
Those bosonic Green's functions are rewritten as 
$ G^{<} = 2 i n(\omega ) {\rm{Im}}  G^{\rm{r}}$ and $ G^{>} = 2 i [1+n(\omega )] {\rm{Im}}  G^{\rm{r}}$,
where $ G^{\rm{r}} $ denotes the retarded Green's function and $ n(\omega )=({\rm{e}}^{\beta \hbar \omega }-1)^{-1}  $ is the Bose-distribution function.
%%%%%%%%%%%%%%%%%%%%%%%%%%%%%%%%%
Introducing $\Delta n(\omega ) \equiv   n_{\rm{R}}(\omega ) - n_{\rm{L}}(\omega ) $, those Green's functions become
$ G_{{\rm{L}}}^{\rm{<}}  G_{{\rm{R}}}^{\rm{>}}   - G_{{\rm{L}}}^{\rm{>}}  G_{{\rm{R}}}^{\rm{<}} 
= 4  {\rm{Im}}  G^{\rm{r}}_{{\rm{L}}}   {\rm{Im}}  G^{\rm{r}}_{{\rm{R}}} \cdot  \Delta n(\omega )   $ and 
$ G_{{\rm{L}}}^{\rm{<}}  G_{{\rm{R}}}^{\rm{>}}   + G_{{\rm{L}}}^{\rm{>}}  G_{{\rm{R}}}^{\rm{<}} 
= - 4  {\rm{Im}}  G^{\rm{r}}_{{\rm{L}}}   {\rm{Im}}  G^{\rm{r}}_{{\rm{R}}} 
[2 {\cal{F}}_0 (\omega ) +   {\rm{coth}}(\beta \hbar \omega /2) \cdot  \Delta n(\omega )] $.
%%%%%%%%%%%%%%%%%%%%%%%%%%%%%%%%%
Thus we obtain Eqs. (\ref{eqn:Im})-(\ref{eqn:SQ}) in the main text.

%%%%%%%%%%%%%%%%%%%%%%%%%%%%%%%%%
 \section{Onsager matrix element}
\label{sec:Appe2}
%%%%%%%%%%%%%%%%%%%%%%%%%%%%%%%%%

In the presence of an applied strong magnetic field $B_{\rm{L(R)}}$, assuming the Zeeman term $ g \mu_{\rm{B}} B_{\rm{L(R)}}$ and expanding $\Delta n(\omega )$ 
within the linear response regime in respect to $\Delta B$ and $\Delta T$, 
i.e. $\Delta n(\omega ) =  - \beta  g \mu _{\rm{B}} {\cal{F}}_0 (\omega ) \Delta B +(\beta /T) {\cal{F}}_1 (\omega ) \Delta T   $,
from Eqs. (\ref{eqn:Im})-(\ref{eqn:SQ}) we obtain the Onsager coefficients Eqs. (\ref{eqn:L11})-(\ref{eqn:Smstar}) in the main text.
%%%%%%%%%%%%%%%%%%%%%%%%%%%%%%%%%%

While being temperature-independent, $  {\rm{Im}}  G^{\rm{r}}_{{\rm{L}}}   {\rm{Im}}  G^{\rm{r}}_{{\rm{R}}}  $ is a function of $B_{\rm{L(R)}}$.
Thereby expanding $  {\rm{Im}}  G^{\rm{r}}_{{\rm{L}}}   {\rm{Im}}  G^{\rm{r}}_{{\rm{R}}}  $ in powers of $\Delta B$, from Eqs. (\ref{eqn:Sm}) and (\ref{eqn:SQ}) we obtain the Onsager coefficients $  {\cal{S}}_{11} $ and ${\cal{S}}_{21}  $ [Eq. (\ref{eqn:Matrix})];
\begin{subequations}
\begin{eqnarray}
%%%%%%%%%%%%%%%%%%%%%%%%%%%%%%%%%%
{\cal{S}}_{11} &=& \beta (g \mu _{\rm{B}})^3 \int \frac{d\omega }{2 \pi}   {\cal{G}}   {\cal{F}}_0  {\rm{coth}}(\beta \hbar \omega /2) \nonumber \\
&-& 2 (g \mu _{\rm{B}})^2 \int \frac{d\omega }{2 \pi}   {\cal{G}}'_{\rm{R}}   {\cal{F}}_0,
 \label{eqn:S11} \\
%%%%%%%%%%%%%%%%%%%%%%%%%%%%%%%%%%
{\cal{S}}_{21} &=&  \beta g \mu _{\rm{B}} \int \frac{d\omega }{2 \pi}   {\cal{G}}   {\cal{F}}_2  {\rm{coth}}(\beta \hbar \omega /2) \nonumber \\
&-& 2  \int \frac{d\omega }{2 \pi}   {\cal{G}}'_{\rm{R}}   {\cal{F}}_2,
 \label{eqn:S21} 
%%%%%%%%%%%%%%%%%%%%%%%%%%%%%%%%%%
\end{eqnarray}
\end{subequations}
where
${\cal{G}}'_{\rm{R}} = {\cal{G}}'_{\rm{R}}(B)  \equiv  4 (J_{\rm{ex}}S)^2   \sum_{{\mathbf{k}}, k_x^{\prime}} 
{\rm{Im}} G_{{\rm{L}}, {\mathbf{k}}, \omega }^{\rm{r}} 
(\partial /\partial B) {\rm{Im}} G_{{\rm{R}}, {\mathbf{k}}^{\prime}, \omega }^{\rm{r}}\mid _{B_{\rm{R}}=B}$.
%%%%%%%%%%%%%%%%%%%%%%%%%%%%%%%%%
Note that due to the ${\cal{G}}'_{\rm{R}}$-term, which arise from the ${\cal{F}}_0 $-term in Eqs. (\ref{eqn:Sm}) and (\ref{eqn:SQ}), 
the ratios $ {{\cal{S}}_{11}}/{L_{11}}  $ and  $ {{\cal{S}}_{21}}/{T L_{22}}  $ do not reduce to the value $ g \mu _{\rm{B}}   $ even in the low-temperature limit.
%%%%%%%%%%%%%%%%%%%%%
However, in the magnonic shot noise regime  $  2{\cal{F}}_0 (\omega )   \ll     {\rm{coth}}(\beta \hbar \omega /2) \mid   \Delta n(\omega ) \mid  $,
the ${\cal{G}}'_{\rm{R}}$-term vanishes and those become 
$ {\cal{S}}_{11} = \beta (g \mu _{\rm{B}})^3 \int ({d\omega }/{2 \pi})   {\cal{G}}   {\cal{F}}_0  {\rm{coth}}(\beta \hbar \omega /2) $,
$ {\cal{S}}_{21} =  \beta g \mu _{\rm{B}} \int ({d\omega }/{2 \pi})   {\cal{G}}   {\cal{F}}_2  {\rm{coth}}(\beta \hbar \omega /2) $,
and finally, each coefficient reduces to
$ {{\cal{S}}_{11}}/{L_{11}}  \stackrel{\rightarrow }{=}   {{\cal{S}}_{21}}/{T L_{22}} \stackrel{\rightarrow }{=}  g \mu _{\rm{B}}$
at low temperatures $ 0 \not=  k_{\rm{B}} T \ll   g \mu_{\rm{B}} B $.
%%%%%%%%%%%%%%%%%%%%%%%%%%%%%%%%%%%%
Thereby even if $  {\rm{Im}}  G^{\rm{r}}_{{\rm{L}}}   {\rm{Im}}  G^{\rm{r}}_{{\rm{R}}}  $ is temperature-dependent as a function of $T_{{\rm{L(R)}}}$ (e.g. temperature-dependent lifetime),
those results in the main text qualitatively remain valid in the magnonic shot noise regime; the Onsager coefficients within the linear response region,
$  L_{ij}    $ and $   {\cal{S}}_{ij}    $ with $  {\cal{S}}_{{\rm{m}}(Q)}^{0 }  $, remain unchanged in the magnonic shot noise regime.
%%%%%%%%%%%%%
We remark that in the low-temperature limit of the magnonic shot noise regime, from Eqs. (\ref{eqn:Im}) and (\ref{eqn:Sm}) we obtain
$ {\cal{S}}_{\rm{m}}/{\cal{I}}_{\rm{m}} \stackrel{\rightarrow }{=}   g \mu _{\rm{B}}  $ in any orders of $\Delta T$ and $\Delta B$.

Focusing on a larger temperature difference $ \Delta T$ (i.e. $ g \mu_{\rm{B}}\mid  \Delta  B \mid  \ll     k_{\rm{B}}\mid  \Delta T \mid $) 
and thereby expanding $\Delta n(\omega )$ again but up to ${\cal{O}}\big((\Delta T)^2\big) $, 
$\Delta n(\omega ) =  - \beta  g \mu _{\rm{B}} {\cal{F}}_0 (\omega ) \Delta B +(\beta /T) {\cal{F}}_1 (\omega ) \Delta T  
+ [-(\beta /T^2) {\cal{F}}_1 (\omega )+(\beta ^2/2 T^2) {\cal{F}}_2 (\omega ) {\rm{coth}}(\beta \hbar \omega /2)] (\Delta T)^2    $,
from Eqs. (\ref{eqn:Im})-(\ref{eqn:SQ}) we obtain the Onsager coefficients for the nonlinear responses in respect to $ \Delta T$ [Eq. (\ref{eqn:Matrix2})];
\begin{subequations}
\begin{eqnarray}
%%%%%%%%%%%%%%%%%%%%%%%%
 L_{13} &=&   - \frac{\beta g \mu _{\rm{B}}}{T^2}   \int \frac{d\omega }{2 \pi}   {\cal{G}} {\cal{F}}_1      \nonumber    \\
            &+&      \frac{\beta^2 g \mu _{\rm{B}}}{2 T^2}  \int \frac{d\omega }{2 \pi}   {\cal{G}}  {\cal{F}}_2  {\rm{coth}}(\beta \hbar \omega /2),
\label{eqn:L130}     \\
%%%%%%%%%%%%%%%%%%%%%%%%
 {{\cal{S}}_{13}} &=&  - \frac{\beta (g \mu _{\rm{B}})^2}{T^2}   \int \frac{d\omega }{2 \pi}   {\cal{G}} {\cal{F}}_1  {\rm{coth}}(\beta \hbar \omega /2)  \nonumber \\
              &+&     \frac{(\beta g \mu _{\rm{B}})^2}{2 T^2}  \int \frac{d\omega }{2 \pi}   {\cal{G}}      {\cal{F}}_2  {\rm{coth}}^2(\beta \hbar \omega /2),  \\
\label{eqn:S130} 
%%%%%%%%%%%%%%%%%%%%%%%%
 L_{23} &=&   - \frac{\beta}{T^2}   \int \frac{d\omega }{2 \pi}   {\cal{G}} {\cal{F}}_2   \nonumber \\
              &+&     \frac{\beta ^2}{2 T^2}  \int \frac{d\omega }{2 \pi}   {\cal{G}}      {\cal{F}}_3  {\rm{coth}}(\beta \hbar \omega /2),
 \label{eqn:L230}   \\
 %%%%%%%%%%%%%%%%%%%%%%%%
 {{\cal{S}}_{23}} &=& - \frac{\beta}{T^2}   \int \frac{d\omega }{2 \pi}   {\cal{G}} {\cal{F}}_3   {\rm{coth}}(\beta \hbar \omega /2)  \nonumber \\
              &+&     \frac{\beta ^2}{2 T^2}  \int \frac{d\omega }{2 \pi}   {\cal{G}}      {\cal{F}}_4 {\rm{coth}}^2(\beta \hbar \omega /2).
\label{eqn:S230}     
\end{eqnarray}
\end{subequations}
In the low-temperature limit $ 0 \not= k_{\rm{B}}T \ll  g \mu_{\rm{B}}  B  $, 
the coefficients for the nonlinear responses, except ${\cal{S}}_{23}$, are described by those for the linear responses;
\begin{subequations}
\begin{eqnarray}
L_{13} &\stackrel{\rightarrow }{=}&  \frac{\beta ^2 g \mu _{\rm{B}}}{4T^2}     {\cal{S}}_{Q}^{0 } 
= \frac{\beta ^2 g \mu _{\rm{B}} }{2}  k_{\rm{B}}  L_{22},
\label{eqn:L13}     \\
%%%%%%%%%%%%%%%%%%%%%%%%
 {{\cal{S}}_{13}} & \stackrel{\rightarrow }{=}& \frac{(\beta g \mu _{\rm{B}})^2 }{4T^2}   {\cal{S}}_{Q}^{0 }
 = \frac{(\beta  g \mu _{\rm{B}})^2 }{2}  k_{\rm{B}}  L_{22},
\label{eqn:S13}     \\
%%%%%%%%%%%%%%%%%%%%%%%%
 L_{23} &=& - \frac{1}{T} L_{22} + \frac{1}{2 k_{\rm{B}}T^2} {\cal{S}}_{22}   
 \stackrel{\rightarrow }{=}  \frac{1}{2 k_{\rm{B}}T^2} {\cal{S}}_{22}.
\label{eqn:L23}       
\end{eqnarray}
\end{subequations}
The coefficients $L_{13}$ and ${{\cal{S}}_{13}}$ are seen to be characterized by the thermal noise ${\cal{S}}_{Q}^{0 } $, and those can be rewritten in terms of the thermal conductance $K$ by Eq. (\ref{eqn:S12K}).

Lastly, we remark that in the main text we have considered the exchange interaction $J_{\rm{ex}}$ which conserves the magnetic momentum, and have calculated the tunneling current and the noise as a function of $J_{\rm{ex}}$ [Eqs. (\ref{eqn:Im})-(\ref{eqn:SQ})];
$  {\cal{I}}_{{\rm{m}}(Q)} = {\cal{I}}_{{\rm{m}}(Q)} (J_{\rm{ex}}) \propto  J_{\rm{ex}}^2$, 
$  {\cal{S}}_{{\rm{m}}(Q)}   =  {\cal{S}}_{{\rm{m}}(Q)}    (J_{\rm{ex}})  \propto    J_{\rm{ex}}^2$.
%%%%%%%%%%%%%%%%%%%%%%%%%%%%%%%%
Replacing the parameter $ J_{\rm{ex}} $ by $ J_{\rm{ex}}^{\pm } \equiv   \sqrt{J_{\rm{ex}}^2  \pm   {J'}_{\rm{ex}}^2  } $,
we obtain the current $ {\cal{I}}_{{\rm{m}}(Q)} (J_{\rm{ex}}^{-})  $ and the noise $  {\cal{S}}_{{\rm{m}}(Q)}    (J_{\rm{ex}}^{+})  $ in the presence of an exchange interaction $J'_{\rm{ex}}$ which does not conserve the magnetic momentum during the tunneling process, see Ref. \cite{noise2018} for details.
The Onsager coefficients [Eqs. (\ref{eqn:L11})-(\ref{eqn:Smstar})] become $L_{ij} = L_{ij}  (J_{\rm{ex}}^{-})$ while $  {\cal{S}}_{ij}=  {\cal{S}}_{ij}   (J_{\rm{ex}}^{+})  $ and  $ {\cal{S}}_{{\rm{m}}(Q)}^{0 } =  {\cal{S}}_{{\rm{m}}(Q)}^{0 }  (J_{\rm{ex}}^{+}) $, accordingly.
%%%%%%%%%%%%%%%%%%%%%%%%%%%%%%%%
Thus it can be seen that while the ratio of $ L_{ij}$ to ${\cal{S}}_{ij}$ or ${\cal{S}}_{{\rm{m}}(Q)}^{0 } $ in the main text is modified, the others 
[i.e., $L_{ij}/L_{i' j'}$ ($i', j' = 1, 2$), ${\cal{S}}_{ij}/{\cal{S}}_{i' j'}$, and ${\cal{S}}_{ij}/{\cal{S}}_{{\rm{m}}(Q)}^{0 }$] remain unchanged even in the presence of the exchange interaction $J'_{\rm{ex}}$;
$ L_{ij}(J_{\rm{ex}}^{-})/{\cal{S}}_{ij}(J_{\rm{ex}}^{+})  \not=  L_{ij}(J_{\rm{ex}})/{\cal{S}}_{ij}(J_{\rm{ex}})  $
and 
$ L_{ij}(J_{\rm{ex}}^{-})/{\cal{S}}_{{\rm{m}}(Q)}^{0 }(J_{\rm{ex}}^{+})  \not=  L_{ij}(J_{\rm{ex}})/{\cal{S}}_{{\rm{m}}(Q)}^{0 }(J_{\rm{ex}}) $
%%%%%%%%%%%%%%%%%%%%%%%%%%%%%%%%%
while
$ L_{ij}(J_{\rm{ex}}^{-})/L_{i' j'}(J_{\rm{ex}}^{-}) =  L_{ij}(J_{\rm{ex}})/L_{i' j'}(J_{\rm{ex}})  $,
${\cal{S}}_{ij}(J_{\rm{ex}}^{+})/{\cal{S}}_{i' j'}(J_{\rm{ex}}^{+})={\cal{S}}_{ij}(J_{\rm{ex}})/{\cal{S}}_{i' j'}(J_{\rm{ex}})  $, and 
${\cal{S}}_{ij}(J_{\rm{ex}}^{+})/{\cal{S}}_{{\rm{m}}(Q)}^{0 }(J_{\rm{ex}}^{+}) = {\cal{S}}_{ij}(J_{\rm{ex}})/{\cal{S}}_{{\rm{m}}(Q)}^{0}(J_{\rm{ex}}) $.

\bibliography{PumpingRef}

\end{document}